\documentclass[conference]{IEEEtran}
\IEEEoverridecommandlockouts
\usepackage{cite}
\usepackage{amsmath,amssymb,amsfonts}
\usepackage{algorithmic}
\usepackage{graphicx}
\usepackage{textcomp}
\usepackage{xcolor}
\usepackage{float}
\usepackage{hyperref}
\def\BibTeX{{\rm B\kern-.05em{\sc i\kern-.025em b}\kern-.08em
 T\kern-.1667em\lower.7ex\hbox{E}\kern-.125emX}}
\def\CLASSOPTIONpaper{letter}%

\begin{document}

\title{Simulation and computational analysis of multiscale graph agent-based tumor model\\
{\footnotesize \textsuperscript{}}
\thanks{}
}

\author{\IEEEauthorblockN{\textsuperscript{} Ghazal Tashakor}
\IEEEauthorblockA{\textit{Department Computer Architecture \& Operating Systems} \\
\textit{Universitat Autònoma de Barcelona}\\
Barcelona, Spain \\
Ghazal.Tashakor@uab.cat}
\and
\IEEEauthorblockN{\textsuperscript{} Remo Suppi}
\IEEEauthorblockA{\textit{Department Computer Architecture \& Operating Systems} \\
\textit{ Universitat Autònoma de Barcelona}\\
Barcelona, Spain\\
Remo.Suppi@uab.cat}%
\thanks{Research supported under contract TIN2017-84875-P, funded by the Agencia Estatal de Investigaci\'on, Spain and the FEDER-UE and partially funded by a research collaboration agreement with the Fundaci\'on Escuelas Universitarias Gimbernat.}
\and

\and

\and

\and

}

\maketitle

\begin{abstract}
This paper deals with the cellular biological network analysis of the tumor-growth model, consisting of multiple spaces and time scales.
In this paper, we present a model in graph simulation using ABM for tumor growth. In particular, we propose a graph agent-based modeling and simulation system in the format of tumor growth scenario for evolving analysis. To manage cellular biological network analysis, we developed a workflow that allows us to estimate the tumor model and the complexity of the evolving behavior in a principled manner. By developing the model using Python, which has enabled us to run the model multiple times (more than what is possible by conventional means) to generate a large amount of data, we have succeeded in getting deep in to the micro-environment of the tumor, employing network analysis.\\
Combining agent-based modeling with graph-based modeling to simulate the structure, dynamics, and functions of complex networks is exclusively important for biological systems with a large number of open parameters, e.g., epidemic models of disease spreading or cancer.
Extracting data from evolutionary directed graphs and a set of centrality algorithms helps us to tackle the problems of pathway analysis and to develop the ability to predict, control, and design the function of metabolisms.
Reproducing and performing complex parametric simulations a known phenomenon at a sufficient level of detail for computational biology could be an impressive achievement for fast analysis purposes in clinics, both on the predictive diagnostic and therapeutic side.
\end{abstract}


\section{Introduction}
Scientific agent-based modeling and simulation requires specific techniques to manage parametric executions and the computational cost of the evolution analysis. It gets more complicated when it is defined as Systems Biology (SB) based on a multiscale nature.
There is a large set of references in SB which reviewed and compared different agent-based modeling tool-kits.
From the scientific community point of view this field still lacks an accepted generic methodology to address multiscale computation. Specifically to optimize the transport of data between sub-models in high-performance computing (HPC) environments which means exchanging large volumes of data. In this case, dedicated data pattern software and high-performance multiscale computing applications would be needed.

This paper presents our work in graph simulation and modeling of an agent-based tumor growth. The most novel changes in this paper on the ground of the other scenarios are obtaining a probabilistic state machine in Python using Mesa [4], a graph-based representation using NetworkX [21] and consideration of multiscale simulation to perform cellular and molecular population dynamics.
Unlike many other tools, NetworkX [21] data analysis features are designed to handle data on a scale relevant to complex problems, and most of the core algorithms rely on extremely fast legacy code for flexible graph representations which a node and edge can be anything.

On the other hand, implementing our tumor agent-based model in Python using Mesa [4] is indeed a prospective approach since it is becoming the language for scientific computing and facilitating the web crawling for direct visualization of every model step. Mesa allows large-scale modeling to create agent-based models using built-in core components such as agent schedulers and spatial grids in parallel computation. 

\section{Background}
A primary tumor model just addressing the vascular growth state depends on differential equations, but In silico [6] refers to computational models of biology and it has many applications.
There are three approaches of In silico to build a cancer model: continuum, discrete and hybrid. Each approach has the characteristics that make it suitable for analyzing specific properties of tumors and tumor cells [8]. However, an adaptive hybrid model which integrates both continuous and discrete based models is the most challenging for simulating a complex system.
A minimal coupling of a vascular tumor dynamics to tumor angiogenic factors through agent-based modeling has pushed the progress of experimental studies during recent years\footnote{ABM model that allows rule definition with great detail and different levels of activation to model the angiogenesis phenomenon.}. It is a big challenge to simulate total process of a complex system such as tumor growth, metastasis, and tumor response treatments mathematically because mathematical modeling is still a simplification of the systems biology and the results require validation [7].
As we can see in some papers such as [8][9], it is apparent that building a bulky model over a range of matrix densities which covers numerous factors in this way for large domain sizes or 3D simulations are restricted by computational and application costs. In [8] they have presented a series of ABMs that are intended to introduce a multiscale architecture\footnote{Multilevel structure of information exchange through which it is defined that how the agents interact in the model.} for representing biomedical knowledge in NetLogo [1] and in conclusion they mentioned that a full-scale ABM implementation is not possible at this time. 
They have pointed out that there is a need to develop and communicate the potential framework that is conceptually robust and allows the evolution of knowledge represented in a computable form.
Authors of [10][11] also suggested focusing on modified methods for analyzing and modeling which scales more with network size using information about edge betweenness to detect community peripheries\footnote{Methods proposed by [10-12] are based on the exchange of information between agents that allows extracting a greater quantity of tumor information and understanding its dynamics}. This suggestion allows scientists to go through a broader investigation of tumor information extraction and tumor-growth dynamics.
Also, it seems using dynamic networks based on a large number of interactive agents make it possible for researchers to carry out more detailed research on inter-cellular network interactions and metastasis in a multiscale model [12].
One of the latest papers is [22] which addresses recent progress and open questions in multiscale modeling.
From their point of view, a well-established methodology is building and maintaining a computer code and proposing a framework which includes theoretical concepts, a multiscale modeling language and an execution environment to solve the interdisciplinary multiscale problems such as spatial scale.
On the other hand, cell populations can be very heterogeneous, so nested effects modeling for single-cell data to simultaneously identify different cellular sub-populations to explain the heterogeneity in a cell population will be necessary. It helps entering the mechanisms of gene regulation and the reconstruction of cell signaling networks [23].

Building a general framework at this time which could model the static and dynamic aspects of the tumor behavior and provide the computational resources is an open application area challenge with many technical barriers overcome in recent years [23]. Perhaps a hybrid graph agent-based simulation and modeling could be an outset to metabolic engineering for applications which have expanded to address problems such as evolution.

It is noteworthy that graphs, in general, are useful in such integrative analysis of data from different sources which naturally required an in-depth integrity and dependency.
\subsection{Modeling and Simulation in NetLogo}
Our initial ABM NetLogo model [2] was designed as a self-organized model that illustrates the growth of a tumor and how it resists chemical treatment. This model in NetLogo which is based on Wilensky's tumor model [1] permits us to change the parameters that affect tumor progression, immune system response, and vascularization.
\begin{figure}
\centering
\includegraphics[width=70mm,scale=0.5,]{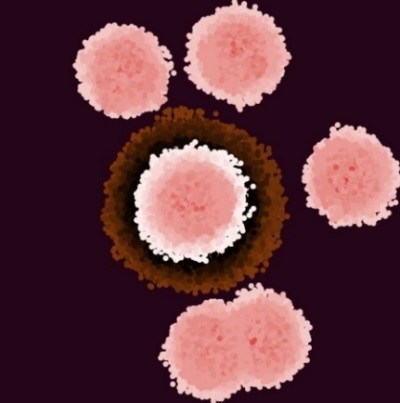}
\caption{Stem cells evolution and metastasis visualization with grow-factor=1.75, apoptosis=low and replication-factor=high (near 200,000 cells in steady state).} \label{fig1}
\end{figure}
Figure 1 shows the steady state of a tumor metastasis visualization with six stem cells and the grow-factor=1.75, replication-factor=high, and apoptosis=low.
As it could be seen, the growth of metastasis is more aggressive and through reducing apoptosis, there is a higher number of cells that do not die, amounting near 200,000 cells (agents) [2].
NetLogo [1] includes the Behavior Space tool that allows the exploration of the model data space using parametric executions in varying settings of the model and for recording the results of each model run.
The main problem of these executions is that the Behavioral Space only supports multithreading, so its performance is limited to the number of cores/threads at the local infrastructure. To solve the problem, we have executed the parametric simulations using our HPC cluster in order to reduce the necessary time to explore a determinate model data space.
It has pointed out that this implementation using NetLogo caused the limitations of the execution environment (Java memory limitations) and loss of performance with a high number of metastasis cells.
Also, it shows this model did not allow capturing in detail interactions between the different parameters the microenvironment level.

\subsection{Static preliminary model in Python}
\begin{figure}
\centering
\includegraphics[width=70mm,scale=0.5]{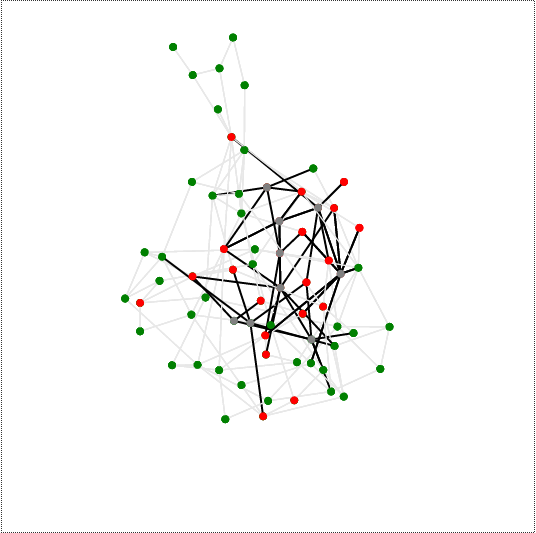}
\caption{Graph visualization for a tumor in three states (normal, dead and inflamed) which has shown in three colors(green, grey and red).} \label{fig2}
\end{figure}

Taking into account the limitations of Netlogo and the different research tendencies to introduce multiscale simulations to represent biomedical knowledge, our research was oriented towards this type of ABM simulations. For this, the environment and modeling were changed to represent all the interactions in a multiscale ABM model. In this sense, Python + Mesa were chosen as a development environment and a graph-based model was selected to represent all the complexity and interactions of the tumor model. Fig.~\ref{fig2} shows the visual form of our second approach [5] for tumor agent-based modeling which tumor cells changed color while they go through state transition.
We have implemented a preliminary tumor agent-based model in graph architecture considering that in life science data analysis such as tumor almost everything is about connections and dependencies. As well a large amount of data makes it difficult for researchers to identify insights or controlling dynamical networks and here the role of graph architecture in displaying complex patterns of interactions between components is significant.
Each agent nested in a single-node and changes in three states under the influence of the neighbor nodes. The process goes on until the tumor agent's volume appears as metastasis.
To simulate a graph network for this model, we selected a random graph to construct the cell interactions and stromal cells behavior within a tumor microenvironment.
Due to the time-dependency of the connections, we developed our graph agent-based model on Erdös-Rényi [24] topology. Erdös-Rényi model takes a number of vertices N and connecting nodes by selecting edges from the (N (N-1)/2) possible edges randomly.
As the same scenario as our NetLogo tumor model, to study and analyze the behavior of Python tumor model under different conditions, we needed to explore the relevant data of the model using parametric executions. For this purpose, we found Mesa [4] which supports multi-agent and multiscale simulations. It is a framework that allows us to make changes to existing ABMs. Also monitoring the data management issues when processing actions happen in parallel seems facile in Mesa since each module runs on the server and turns a model state into JSON data.
Graph visualization makes large amounts of data more accessible and easier to read as we can see in Fig.~\ref{fig2}.
The interactive visualization in Mesa helps us to identify insights and generate value from connected data. The visualization of the model is a network of nodes that shows the distribution of agents and their links. A scheduler (time module) activates agents and stores their locations and updates the network. The total operation time is directly related to the number of steps necessary to deploy all the agents.
\section{A new multiscale graph agent-based model of tumor-growth}
Large multiscale experimental modeling and simulation causes an accumulation of data which reflect the possible infinite divers of interactions in cellular biological networks. Accordingly, there is a great need for computational methods and computer tools to manage, query and analysis of these experiments. In the most abstract level, cellular biological networks represent as mathematical graphs because metabolic networks generally require complex representations which have made it possible to investigate the topology and functions of these kinds of networks.
By using graph-theoretical concepts, predicting the dynamical properties of deep layers may suggest new biological hypotheses. Functional modules across different data sources will be essential in understanding the behavior of the system on a large scale.
Since agent-based modeling became an alternative and potentially a more appropriate form of mathematics to define for a computational system, ABM researchers use model analyzing sample distributions to record real-world network outcomes and summarize the theoretical concepts. In the end, the results of these analyses also can be beneficial on the biological goals of the study.
The development and clinical implementation for tumor growth behaviors have become a priority these days, and it requires the analysis of large multiscale data from cell populations to identify features and parameters which predict tumor behavior.
Our static preliminary model in Python still was a limited scale model because of using the Erdös-Rényi graph. To advance the initial idea, we have developed a computational workflow for simulating a multiscale tumor model. The graph-based methodology nested in agent-based modeling aids us to exploit evolving analysis more accurate. Mapping agents to the nodes in the graph-structure model coordinates the assignments of values to their variables in such a way that maximizes their aggregation. Agents work as states, locations or even sometimes as controls of all the variables that map to the nodes.
\subsection{A workflow for simulating tumor model and evolving analysis}\label{AA}
\begin{figure}[H]
\includegraphics[width=90mm,scale=0.5]{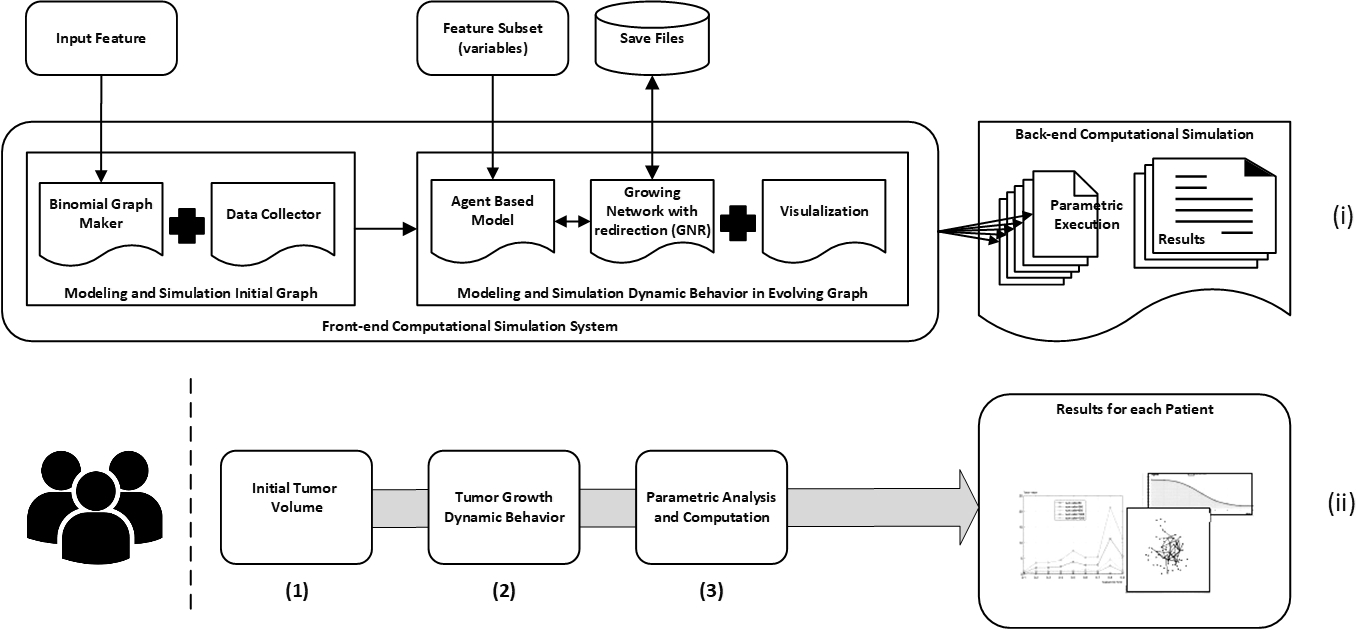}
\caption{Graph-based modeling and simulation system in the format of tumor growth scenario} \label{fig3}
\end{figure}

This section aims to illustrate an evolving analysis of tumor growth in different patients. Let's assume that an oncologist needs growing analysis of cellular interplay for a patient with newly diagnosed cancer disease. Therefore the basic level of the tumor must be characterized for future prediction of the possible growth behaviors. In principle, we need intervention by agent-based modeling to set the initial experimental conditions.
Fig.~\ref{fig3} introduces the workflow of our computational simulation system for modeling a scenario of tumor growth and preparing data from different scales and stages of its evolving behavior. As can be seen at the part (ii) in Fig.~\ref{fig3}, we have defined three steps for the scenario of the workflow.
The first step is simulating an initial tumor by setting up initial features to create an initial graph model. Collecting data at the end of each step helps the oncologist to reuse the data whenever he needs again.

The second step is forwarding the initial graph to the growing module with redirection possibilities in tumor growth. The third step is feeding the growing tumor by changing subset features of angiogenic switch.

The last two steps of the scenario are implemented as a growing network with redirection (GNR) [21] nested in an ABM model. 
The subset features of the angiogenic switch define as a state transition based on probabilistic state machine (explained in next section). We considered that the transient states probability adopted in the subset features are necessarily valued between 0 and 1.
Finally, in the Back-end computational part of the workflow, we designed a data visualization tool for the oncologists.

We have deployed two complementary graph-driven methods for analyzing and estimating probable growing network patterns. The selected methods are presented in the context of network analysis in Python using Mesa [4] and NetworkX [21] packages. The amount of data which is extracted from the methods address some problems in cell biology. 

We chose a fast binomial graph generator on Erdös-Rényi topology for initialization, so we used our static preliminary model as an initial input. To simulate our growing network nested in the agent-based model, we used a growing network with redirection (GNR) graph with probability (p) for adding nodes one at a time with a link to the initial nodes. In this graph, a target node is a node where a new link attached. Target nodes are selected randomly following a uniform distribution. We set the redirection probability (p) which is shown by the equation in (1) like that gives us a new pattern of tumor growth every time.

Based on [13], we used the following equation in (1) to calculate the probability (p) of tumor growth. It contains five biological parameters which are known as cancer driver: the number of divisions (d), the number of stem cells (N), the number of critical rate-limiting pathway driver mutations (k), and the mutation rate (u).
\noindent 
 \begin{equation}
 p = 1 - (1 - (1 - (1 - u)^d)^k)^N
 \end{equation}
Subset features get in the computational simulation system through agents as state transition probabilities to change the state of the cells in the growing network. These parameters could be selected by the oncologist or any other user of the system through the interactive visualization form.

At the agent-based model, tumor cells are affected, inflamed and turn quiescent. Based on these key factors we have simulated the tumor growth behavior and measurements such as the tumor volume, density and also we calculate a number of dead, inflamed and tumor-derived cells. 

\subsection{The probabilistic state machine of tumor growth}\label{AB}
Using probabilistic automaton (PA) in computational biology can be a useful aspect of tracking a natural problem.

\begin{figure}[H]
\includegraphics[width=90mm,scale=0.5]{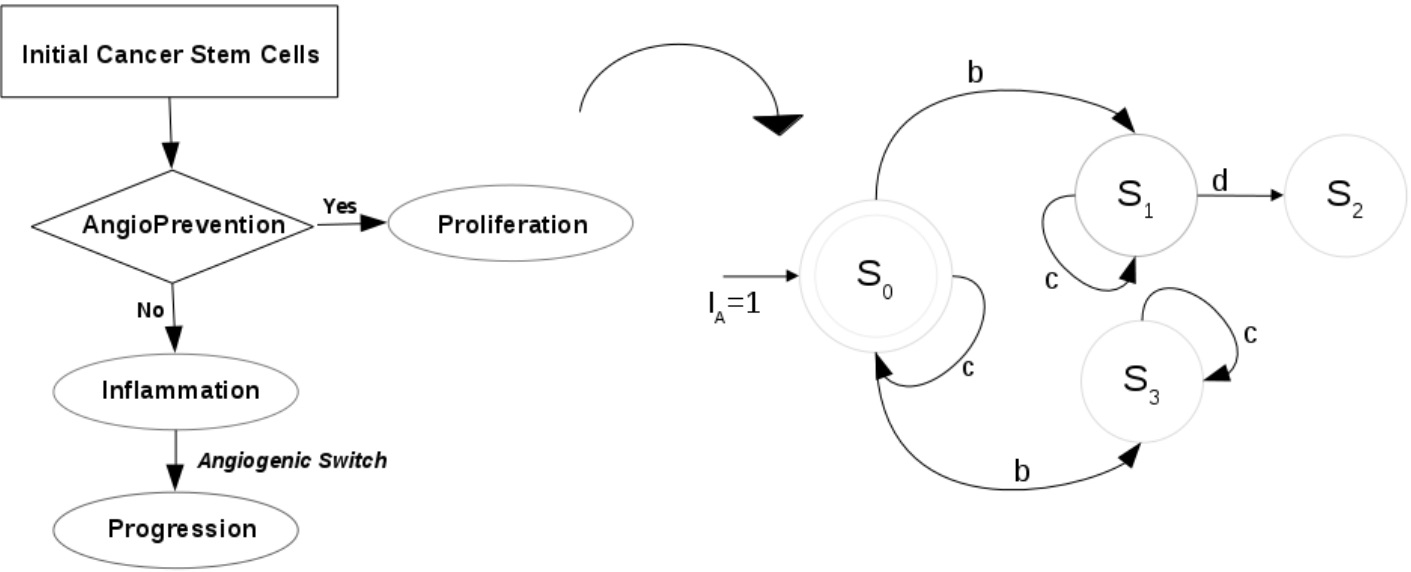}
\caption{Flowchart representation of our tumor behavior in a PFA model } \label{fig4}
\end{figure}
Probabilistic finite-state automaton (PFA) during the past years was applied to the model and generate distributions over sets of possible infinite terms and trees. Typically, PFA is represented as directed labeled graphs[15][16].

The fundamental biological aspect of the probabilistic model of our tumor growth comes from the acute inflammation based upon the key factors involved such as an angiogenic switch. Tumor angiogenesis is critical for tumor growth and maintenance [14][20]. 

Fig.~\ref{fig4} shows the metabolic flowchart of this aspect and how we have created our PFA model based on the flowchart.
The strategy begins with initial identification of a minor population of cells with the characteristics of “tumor-initiating cancer stem cells” and they will be assumed inflamed or dead under the influence of angiogenic switch factors. 

The threshold of angioprevention $K$ is compared with the assessment values of transition probability $P_A$ which is selected by oncologist interactively. The result of the comparison works as a trigger to change the state of the cells from their current state $S_0$ to the proliferation state $S_1$ or the inflammation state $S_3$. Afterward, under the influence of changing angiogenic switch values $\Omega$, the inflammation state $S_1$ may turn to the progression state $S_2$ and metastasis can happen.

\noindent
\begin{equation}
\sum_{s \in S_A}^{}I_A(S)=1
\end{equation}

\begin{equation}
F_A(s)+ \sum_{s'\in S_A, b\in \Omega}^{} P_A(s,b.s')=1 \quad \textrm{where} \quad \forall \ P_s > K_i
\end{equation}
From the probabilistic point of view in equation 2 and 3, a finite generic states $S_A$ is transitioned under the influence of angiogenic probabilities defined by alphabets $\Omega$. $K$ is a threshold drawn from a uniform [0,1] distribution. If the transition probability $P_A$ is greater than $K$, the current state assumed to be extended with $P_A(s,b.s')$. In the End, initial-state probability $I_A$, final state probability $F_A$ and transition probabilities $P_A$ are considered as total and \textit{PFA} definition will be a tuple of below functions.

\subsection{Results of Parametric Execution}\label{AC}
Human-tumor-derived cell lines contain common and different transforming genomic profiles which is essential for a comprehensive understanding of tumorigenesis, and for identifying the earliest events in tumor evolution [17].

\begin{figure}[H]
\centering
\includegraphics[width=90mm,scale=0.5]{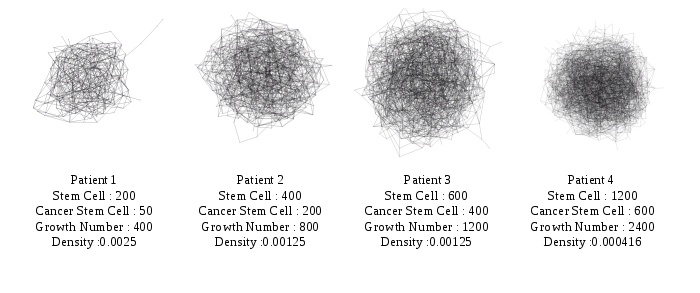}
\caption{Initial tumor graph to set up four baselines (Symbol of four patients) } \label{fig5}
\end{figure}

We assumed four different parametric baseline executions for monitoring tumor-growth model in four different patients as it can be seen in Fig.~\ref{fig5}. For each baseline, it is considered three repetitions as growth patterns to be able to extract data from the growing network in different redirection patterns. The goal is simulating the dynamic behavior of tumor-growth.

The first baseline set to an initial state of 200 stem cells and increase to 400 under the influence of 50 cancer stem cells. The second baseline set to an initial state of 400 stem cells and increase to 800 under the influence of 200 cancer stem cells. The third baseline set to an initial state of 600 stem cells and increase to 1200 under the influence of 400 cancer stem cells. Moreover, the forth baseline set to an initial state of 1200 stem cells and increase to 2400 under the influence of 650 cancer stem cells.

The configuration management of these four baselines is established for analyzing tumor density variations, tumor-derived cells into cancer redirection and identifying genomic profiles of those essential cells. Significant revision regarding graph patterns configuration is extracted from neighborhood analysis and graph centrality methods such as centrality closeness or betweenness [19] which quantify the number of times a node acts as a bridge along the shortest path between two other nodes. These methods were introduced as a measure for quantifying the control of a human on the communication between other humans in a social network by Linton Freeman. 
In this conception, vertices which have a high probability to occur on a randomly chosen shortest path between two randomly chosen vertices have a high betweenness [21].
We used this concept to reconstruct the tumor-derived cell lines and produce their profiles by extracting data from our experiments and comparing the results of the baselines in Table~\ref{tab2} to Table~\ref{tab5}.

As it could be seen, the most aggressive growth pattern belongs to the Table~\ref{tab4} about patient3 for as-much as the average value of the essential genomic profile of tumor patterns and the number of tumor-derived cells are higher than the other tumors.

\begin{figure}[H]
\centering
\includegraphics[width=90mm,scale=0.5]{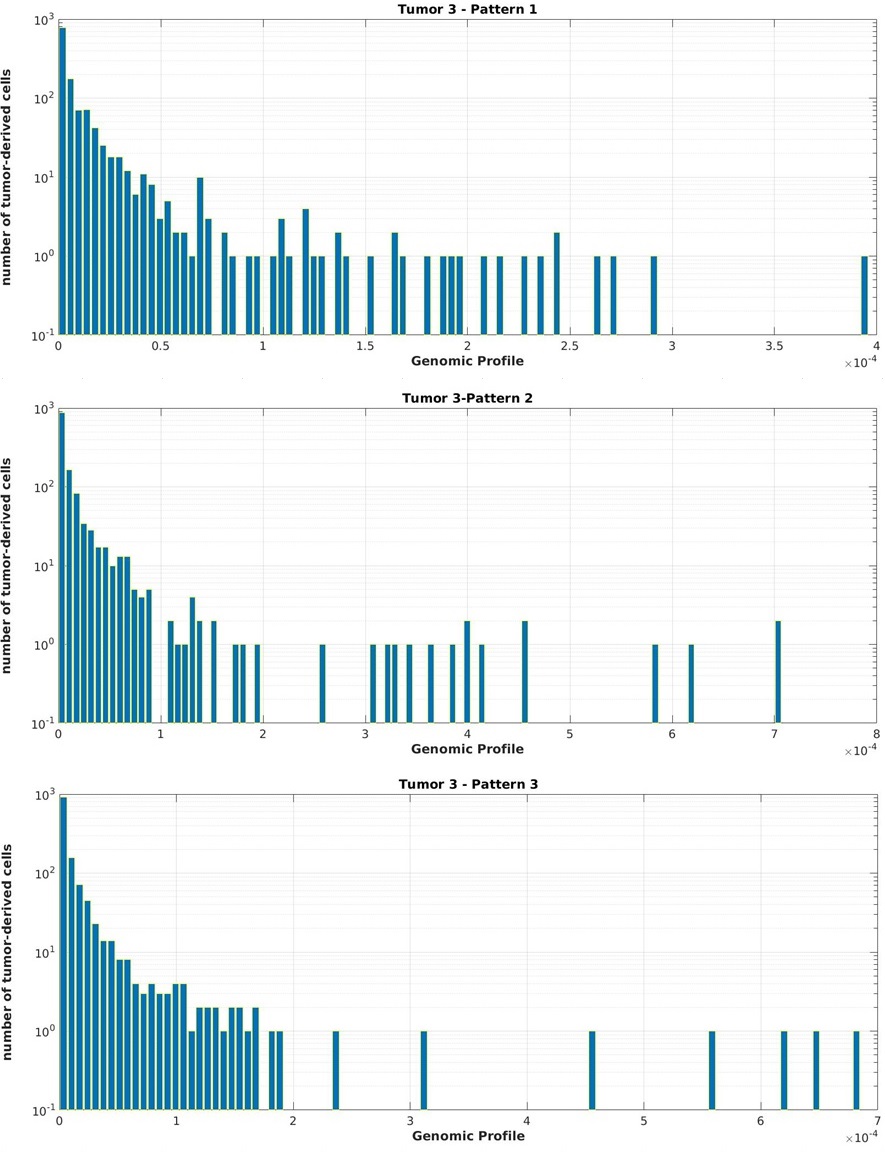}
\caption{Genome profile variation of tumor-derived cells distribution in tumor number three} \label{fig6}
\end{figure}

\begin{figure}[H]
\centering
\includegraphics[width=90mm,scale=0.5]{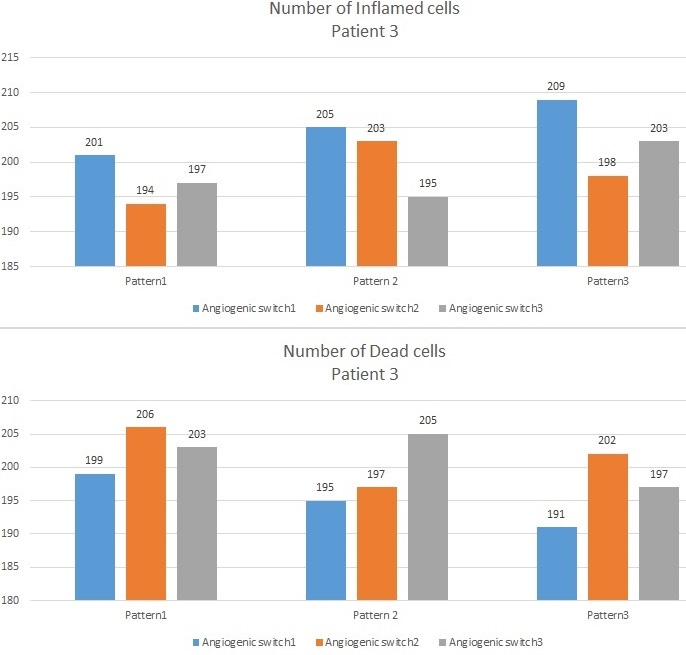}
\caption{The result of the change scale of inflamed and dead cells of tumor number 3} \label{fig7}
\end{figure}

\begin{table}[H]
\centering
\caption{Tumor-derived cell ID and Genomic Profile for Patient 1}\label{tab2}
\begin{tabular}{|c|c|c|c|}
\hline
Initial tumor(Patient1) & GP1 & GP2 & GP3 \\
\hline
tumor-derived cell ID & {\mdseries 10} & {\mdseries 4} & {\mdseries 18}\\
\hline
Essential Genomic Profile & {\mdseries 2.19E-03} & {\mdseries 4.70E-03} & {\mdseries 3.34E-03}\\
\hline
\end{tabular}
\end{table}
\begin{table}[H]
\centering
\caption{Tumor-derived cell ID and Genomic Profile for Patient 2}\label{tab3}
\begin{tabular}{|c|c|c|c|}
\hline
Initial tumor(Patient2) & GP1 & GP2 & GP3\\
\hline
tumor-derived cell ID & {\mdseries 6} & {\mdseries 12} & {\mdseries 6}\\
\hline
Essential Genomic Profile& {\mdseries 1.34E-03} & {\mdseries 8.94E-04} & {\mdseries 1.50E-03}\\
\hline
\end{tabular}
\end{table}
\begin{table}[H]
\centering
\caption{Tumor-derived cells ID and Genomic Profile for Patient 3}\label{tab4}
\begin{tabular}{|c|c|c|c|}
\hline
Initial tumor(Patient3) & GP1 & GP2 & GP3\\
\hline
tumor-derived cell ID & {\mdseries1} & {\mdseries 17 and 10} & {\mdseries 5}\\
\hline
Essential Genomic Profile& {\mdseries 3.96E-04} & {\mdseries 7.07E-04} & {\mdseries 6.85E-04}\\
\hline
\end{tabular}
\end{table}
\begin{table}[H]
\centering
\caption{Tumor-derived cells ID and Genomic Profile for Patient 4}\label{tab5}
\begin{tabular}{|c|c|c|c|}
\hline
Initial tumor(Patient4) & GP1 & GP2 & GP3\\
\hline
tumor-derived cell ID & {\mdseries 1} & {\mdseries 6} & {\mdseries 1}\\
\hline
Essential Genomic Profile& {\mdseries 2.70E-04} & {\mdseries 3.70E-04} & {\mdseries 2.13E-04}\\
\hline
\end{tabular}
\end{table}
Fig.~\ref{fig6} is a visual representation of genome profile variation of tumor-derived cells distribution in tumor number three. We selected this tumor because it behaved more aggressive than the others especially because it grows very radical in its growth patterns number two and three.
Cell lines serve as models to study cancer biology and to connect genomic variation to angiogenic responses. This modeling can aid in understanding different tumor behavior. The tumor-derived cell distribution results are significant for molecular and cell lines study.
Molecular sub-typing could be done based on gene expression patterns and it helps for tumor-derived cell classification [24]. Accordingly, the assortment of classes in tumor-derived cell distribution is our future work.
By computing the ratio of the dead cells to the inflamed cells, we have also been able to demonstrate different tumor growth behavior upon the effective laboratory condition from the angiogenic switch.
Fig.~\ref{fig7} illustrates the change scale of inflammatory in tumor3 based on different angiogenic key factors determinate in Table~\ref{tab1}. There is evidence in the chart that angiogenesis and inflammation are mutually dependent. 
\begin{table}[H]
\centering
\caption{Angiogenic switch key factors as transition probabilities}\label{tab1}
\begin{tabular}{|c|c|c|c|}
\hline
PA Values & ASW1 & ASW2 &ASW3\\
\hline
AngioPrevention & 0.4 & 0.6 & 0.4 \\
Angiogenesis & 0.6 & 0.4 & 0.6 \\
Quiescent & 0.2 & 0.2 & 0.8\\
\hline
\end{tabular}
\end{table}

Also, stem cell quiescence is a way to control the inflammation in the tumor microenvironment. Increasing angiogenic value in angiogenic switch 1 causes inflammatory reactions and raise the number of inflamed cells. Targeting inflammation by using angioprevention and stop cancer cells from proliferating helps to decrease the number of inflamed cells as it is shown in angiogenic switch 2 and 3 of the chart.

\subsection{Conclusions and Future work}
In this paper, we simulated and developed a multiscale graph agent-based model. The model uses for extracting and analyzing data from the cellular network of a tumor while growing.
The extracted information from the hybrid simulation with transient probabilities and variable angiogenesis key factors to target emulating the dynamic behavior of tumor-growth seems interesting to oncologists and scientist since they can study the probable predictive power of pathways in the cellular network of the tumor. 
Migrating from NetLogo to Python using Mesa and NetworkX was a successful strategy since the Python framework permits us to develop faster and deeper into the details in multistage and multiscale modeling. 
Presenting the mathematical and biological form of the tumor-growth model in the format of graph agent-based model using a probabilistic state machine and the identification based on genomic profiling idea could be validated in the future. Also, this idea will allow us to integrate our model to an alternative approach in discovering similarly or densely connected sub-graphs of nodes[18]. This approach is an imitation of the metastasis complication. 

\end{document}